\documentclass[preprint,showpacs,draft,prb,endfloats]{revtex4}
\usepackage{amsmath}
\usepackage{amssymb}
\usepackage{graphicx}
\usepackage{dcolumn}
\usepackage{bm}

\begin{document}

\preprint{APS/123-QED}

\title{Periodicity of the giant vortex states in mesoscopic superconducting rings}

\author{Hu Zhao}
\email{hzh9@bnu.edu.cn}
\author{Shi-jie Yang}
\author{Shiping Feng}
\affiliation{Department of Physics,Beijing Normal
University,Beijing 100875, China
}%

\date{\today}

\begin{abstract}

The giant vortex states of a multiply connected superconductor,
with radius comparable to the penetration depth and the coherence
length, are theoretically investigated based on the nonlinear
Ginzburg-Landau theory, in which the induced magnetic field by the
super-currents is accurately taken into account. The solutions of
Ginzburg-Landau equations are found to be actually independent of
the angular momentum L in a gauge invariant point of view,
provided that the hole is in the center. Different cases with the
paramagnetic current, the diamagnetic current, and the coexistence
of the above two, have been studied numerically. The
interpretation of the L-independent solutions of Ginzburg-Landau
equations is given based on the same principle of Aharonov-Bohm
effect, and could be observed by Little-Parks like oscillations
near the phase boundary.

\end{abstract}



\maketitle

\section{\label{sec:introduction} Introduction}

In 1962, Little and Parks\cite{Little:prl9(9)} demonstrated that
in multiply connected system, the superconducting transition
temperature $T_{c}$ is a periodic function of the applied magnetic
field $H$ in the axial direction of a superconducting thin loop.
The oscillations of $T_{c}(H)$ with a period of the flux quantum
$\Phi_{0}=hc/2e$ are a consequence of the fluxoid quantization
constraint which was first predicted by F. London.\cite{London:su}
The free energy of the superconducting state is periodic in the
unit of flux quanta, while the free energy of normal state is
almost independent of the flux. Therefore the transition
temperature should be a periodic function of the enclosed flux.

Recent advancements in micro-fabrication and technique of Hall
magnetometry \cite{Geim:apl} make it possible to detect the
properties of superconducting samples with sizes comparable to the
coherence length $\xi $ and the London penetration depth $\lambda$
at the temperature well below $T_{c}$ . The superconducting states
in such a mesoscopic system reveal unique properties, which depend
on the size and geometry of the sample. Such confinement effects
have been clearly demonstrated experimentally in nanosized
superconducting square loops\cite{Moshchalkov:nature1},
disks\cite{Geim:nature1} and rings.\cite{Zhang:prb55} In the last
decade, the properties of mesoscopic superconductors have drawn
growing attention, both experimentally and theoretically(see,
e.g., Refs. 7-15).

In particular, as a doubly connected topological sample,
superconducting ring like structures have also been studied
extensively in the past years\cite{Davidovic:prl76,Meyers:prb68}.
In contrast to bulk superconductivity, the symmetry of the sample
play an important role when the sample scales become comparable
with $\xi$ (or $\lambda$). The effective Ginzburg-Landau(GL)
parameter $\kappa= \lambda/\xi$ is significantly increased, and
the magnetic response of a type I superconductor even presents a
shape of a type II superconductor. On the basis of GL theory,
Berger and Rubinstein predicted that there is a surface in a
nonuniform mesoscopic superconducting loop, where the order
parameter vanishes due to the existence of the "singly connected
state". \cite {Berger:prl75} Using the linearized GL equation,
Bruyndoncx {\it et al.} studied superconducting loops of finite
width and found an dimensional crossover from different behaviors
of the phase boundary as the geometry of the loop
changes.\cite{Bruyndoncx:prb60} Taking the induced magnetic fields
into account, Baelus {\it et al.}\cite{Baelus:prb61} investigated
vortex states in a thin superconducting disk with a hole and
noticed a transition from the giant vortex state to the
multivortex state arises when the size of the disk becomes large
enough. They also showed that the multi vortex state is stable
when circular symmetry is broken.

Recently, Pedersen {\it et al.} experimentally investigated the
flux quantization of a mesoscopic loop with a $\mu$ Hall
magnetometer. The magnetic field intensity periodicity observed in
the magnetization measurements indicates a gradual transition from
large flux avalanches to single flux jumps as the external
magnetic field intensity is increased. Such a sub-flux quantum
shift was interpreted as a consequence of a giant vortex state
nucleating towards either the inner or the outer boundary of the
loop.\cite{Pedersen:prb64} Their observation leads to a new
interest in the phase transition in superconducting
loops\cite{Vodolazov:prb66a,Vodolazov:prb66b, Berger:prb67}. At
low magnetic field, multiple flux jumps and irreversible behavior
were also observed by Vodolazov {\it et,
al.}\cite{Vodolazov:prb67} in thin Al superconducting rings with
sufficiently large radii. The also showed the possibility of the
existence of flux avalanches by numerical calculations based on
the time dependent GL model. The nucleation of superconductivity
in rings has been studied both experimentally and theoretically by
Morelle {\it et al.}\cite{Morelle:prb70}, A transition from a
one-dimensional to a two-dimensional regime is seen when the
magnetic field with small holes is increased. The phase
transitions due to the superconducting vortices enter into the
sample were also reported by Bourgeois {\it et al.} in mesoscopic
square loops\cite{Bourgeois: prl94}. The measurements of specific
heat as a function of external magnetic field exhibit oscillations
with changing periodicity, depending on both the temperature and
applied flux.

There are three kinds of vortex states in mesoscopic systems, the
giant vortex state, the multi vortex state and the ring-like
vortex state. In terms of thermodynamical stability, the
multi-vortex states and giant vortex states can be stable, while
ring-like vortices in superconductors seem to be unstable \cite
{Stenuit:physica00, Milosevic:prb02}. In present work, we confined
our study in the giant vortex states since the mesoscopic
superconducting ring has more surfaces compared to the disk with
same size. The confinement effects from the boundaries are
dominating and it will impose a circular symmetry on
superconducting order parameter. Moreover, recent experimental
observations also indicate giant vortex states nucleating in
mesoscopic superconductors with cylindrical
symmetry\cite{Pedersen:prb64, Kanda:prl93}. For a giant vortex
state, the free energy of a type I superconducting ring has been
studied theoretically\cite{Baelus:prb61} in terms of external
magnetic field and the winding number L. The vorticity of the
ground state changes from $L$ to $L+1$ at some values as the
external magnetic field increases. However, their numerical
results seem to show that the properties of giant vortex states
are related to the winding number L. In the present work, we
theoretically investigate the giant vortex states of a multiply
connected superconductor in terms of the nonlinear GL theory. We
show that the winding number L of a giant vortex state is a tuned
function of the flux inside the hole of a ring, and that all the
information of giant vortex states can be obtained with a defined
winding number L. We also show that the vector potential-tuned
giant vortex states have periodicity in the units of $\Phi_{0}$,
as demonstrated in Little-Parks experiment.

This paper is organized as follows: In section II, we describe the
model of the mesoscopic superconducting ring based on the
two-dimensional GL equations. This approach is suitable in two
cases: (i) for a ring made from a thin film and (ii) for a high
cylinder considered in its cross-section far from the bases. We
present a simple proof to show that the Gibbs free energy of a
giant vortex state is independent of the vorticity L. In section
III, we investigate the giant vortex states numerically and obtain
the local magnetic field, the Cooper pair density and the current
density. We consider different cases with diamagnetic response,
paramagnetic response and the combination of the above twos. We
also show how the quantum phase tuned by the flux inside the hole.
In section IV, we give further discussion and conclusions.

\section{The GL equations and the L-independence of Gibbs free energy}

In the present paper, we consider a two-dimensional
superconducting ring, with internal radius $r$ and external radius
$R$. The applied magnetic field is along the axial direction and
has a uniform value$H_{0}$. When the size of the superconducting
ring falls in a mesoscopic regime, where the surface effects are
of the same order of magnitude as the bulk effects, the
confinement effects from the boundaries are dominating and impose
a circular symmetry on the superconducting order parameter and the
local magnetic field. Hence the order parameter is expected to be
given by
\begin{equation}
\psi(\rho,\theta)=f(\rho)e^{i\psi(\rho,\theta)},
\end{equation}
and the magnetic field has the form $ \vec H = \vec{\nabla}\times
{\vec{A}} = H(\rho){\vec {e_z}}$.

We can always choose appropriate gauge so that the vector
potential can be written as,

\begin{eqnarray}
\: \vec A &=& A(\rho){\vec {e_{\theta}}}
\end{eqnarray}

In order to simplify the numerical calculation, we will make the
formulas dimensionless by measuring the order parameter $\Psi $ in units of $\sqrt{%
\displaystyle{\frac{-\alpha }{\beta }}}$, lengths in units of $\sqrt{2}{%
\lambda }$, the magnetic field in units of $\displaystyle{\frac{{\phi }_{0}}{%
4\pi {\lambda }^{2}}}$, the Gibbs free energy in units of $\displaystyle{%
\frac{{H_{c}}^{2}}{4\pi }}$, and the vector potential in units of $%
\displaystyle{\frac{{\phi }_{0}}{2\sqrt{2}\pi \lambda }}$, where ${\phi }%
_{0}=\displaystyle{\frac{hc}{2e}}$ and $H_{c}=\displaystyle{\frac{{%
\phi }_{0}}{4\pi {\lambda }^{2}}}\sqrt{2}{\kappa }$. Then the Gibbs free energy $g=g_{s}-g_{n}$ is given by%

\begin{equation}
g=\int [{\frac{1}{2}}(H-H_{0})^{2}+{\kappa }^{2}(1-|\Psi
|^{2})^{2}-{\kappa}^{2}+|(\vec{\nabla}-i\vec{A})\Psi |^{2}]dV.
\end{equation}

By minimizing the Gibbs free energy, GL equations can be obtained
in the following form:

\begin{equation}
2{\kappa }^{2}\Psi(1-|\Psi |^{2})=(\vec{\nabla} -i\vec{A})^{2}\Psi
\end{equation}

\begin{equation}
\vec{\nabla}\times\vec{\nabla}\times\vec{A}=(\Psi^{*}
\vec{\nabla}\Psi - \Psi\vec{\nabla}\Psi^{*})-2|\Psi|^{2} \vec{A}.
\end{equation}

Substituting the expressions of the order parameter and the vector
potential into GL equations, we have

\begin{eqnarray}
2 \displaystyle{\frac {\partial f}{\partial \rho}} {\frac
{\partial \phi}{\partial {\rho}}}+{\frac {f}{\rho}}{\frac
{\partial \phi}{\partial {\rho}}} +f{\frac {{\partial}^{2}
\phi}{\partial {\rho}^{2}}}+ {\frac {f}{{\rho}^2}}{\frac
{{\partial}^{2} \phi}{\partial {\theta}^{2}}} &=& 0
\end{eqnarray}

\begin{eqnarray}
\displaystyle({\frac {A}{{\rho}^2}}-{\frac
{1}{\rho}}{\frac{\partial A}{\partial \rho}}-{\frac{{\partial}^{2}
A}{\partial {\rho}^{2}}}){\vec {e_{\theta}}} &=&
2f^{2}({\frac{1}{\rho}}{\frac{
\partial \phi}{\partial \theta}}-A){\vec {e_{\theta}}}
+ 2f^{2}{\frac{\partial \phi}{\partial \rho}} {\vec {e_{\rho}}}
\end{eqnarray}

The ${\vec {e_{\rho}}}$ term in Equation (7) has to be vanished,
then we have ${\frac{\partial \phi}{\partial \rho}}=0$.
Substituting this equation back into Equation(5), we find the
phase of order parameter satisfies

\begin{eqnarray}
\displaystyle {\frac {{\partial}^{2} \phi}{\partial {\theta}^{2}}}
&=& 0
\end{eqnarray}

Thus it is convenient to assume $\psi(r)=f(r)e^{iL\theta}$, where
L is called as the winding number of the vortex. When the
superconductor is described by such a circular symmetric order
parameter, it is said to be in giant vortex state. GL equations
turn into a set of second-order ordinary differential equations,

\begin{equation}
\displaystyle{\frac{d^{2}f}{d{\rho }^{2}}}+{\frac{1}{\rho
}}{\frac{df}{d\rho }}-({\frac{L}{\rho }}-A)^{2}f+2{\kappa
}^{2}f(1-f^{2})=0~,
\end{equation}

\begin{equation}
\displaystyle{\frac{1}{d{\rho }}}({\frac {1}{\rho}} {\frac
{d({\rho}A)}{d{\rho}}})+2({\frac{L}{\rho }}-A)f^{2}=0.
\end{equation}

The boundary conditions assure that no current passes through the surface, which is
\begin{equation}
\displaystyle{\frac{df}{d\rho }}|_{_{\rho =R}}=
\displaystyle{\frac{df}{d\rho }}{|}_{\rho =r}=0,
\end{equation}

The Gibbs free energy can be written as $g=\int 2\pi \rho Gd\rho
$, which depends only on the value of the kernel $\rho G$ in the
expression
\begin{eqnarray}
\rho G &=&\displaystyle{\frac{\rho}{2}}(A+\rho {\frac{dA}{d\rho
}}-H_{0})^{2}+ \rho {\kappa }^{2}(1-f^{2})^{2}-\rho {\kappa}^{2}
\nonumber \\
&&+({\frac{df}{d\rho }})^{2}\rho +{\frac{f^{2}}{\rho }}({\frac
{L}{\rho}}-A)^{2}.
\end{eqnarray}

Notice that in Equations(9),(10) and (12), the winding number
choice L can fix the gauge choice for the vector potential so that
\begin{equation}
A(L+1,\rho)=A(L,\rho)+1/{\rho}.
\end{equation}

It is straightforward to find that both Gibbs free energy and GL
equations are gauge invariant as long as we choose the winding
number L and the vector potential $\vec A$ satisfying
Equation(13). Consequently, with different winding number L, GL
equations have same solution for the local magnetic field $\vec
H(\rho)$, Copper pair density $f(\rho)$ and current density
$J(\rho)$. It shall be noted that when the winding number changes
with gauge transformation Equation(13), the vector potential will
change respectively, so that although all the observable physical
properties of the superconductor won't change with winding number,
the flux around the superconductor do change. From the gauge
transformation Equation(13),  it can be predicted that the
mesoscopic superconducting ring has the possibility to hold rather
strong magnetic field inside its hole. Actually, this
gauge-invariant solution has similar interpretation as the
well-known Little-Parks experiment. If the winding number and the
flux change together through a gauge invariant transformation, the
physical properties of the thin-walled superconductor does not
change.

\section{Numerical results of non-linear GL equations }

The non-linear GL equations (9-10) can be represented as four
first-order ordinary differential equations, thus four boundary
conditions are required. Apparently the boundary conditions
equation(11) is not sufficient. We set the value of the vector
potential at the external boundary of the ring since the real
magnetic field can not be obtained experimentally although the
applied magnetic field is known. We apply the \textsl{\
}``shooting method'' \cite{Press:book1}, in which trial
integrations\ are ``launched''\textsl{\ }that satisfy the boundary
condition at one endpoint. The discrepancies from the desired
boundary condition at
the other endpoint are used to adjust the starting conditions, until the%
\textsl{\ }boundary conditions at both endpoints are ultimately
satisfied. The relative\textsl{\ }accuracy is set to be $10^{-6}$
for the numerical calculation.

As it has been discussed in the previous section, the full GL
equations can be obtained with a fixed winding number such as
$L=0$. Three kinds of magnetic responses can be obtained
numerically as we change the boundary condition of vector
potential. In Figure.1 and Figure.2, we show that in the case of
the diamagnetic response, Copper pair density decreases and
magnetic field increases when the radius becomes larger. Since the
magnetic filed at the external boundary are much larger than that
at the internal boundary, the super-current induces an negative
magnetic field to keep the applied magnetic field out of the
superconducting ring from the external boundary so that the part
closer to the inner side of the ring always have larger Cooper
pair density. In the case of paramagnetic response, the situation
will be opposite. The magnetic field at the internal boundary are
much larger. The Cooper pair density increases and the magnetic
field decreases when the radius becomes larger. The super-current
induces a positive magnetic filed in order to keep the applied
field out of the internal boundary. In the case of combination of
both effects, the magnetic field first decreases until it reaches
a minimum  and then it increases with the increase of the radius,
the super-current keeps the magnetic field out of the ring from
both boundaries. The total magnetic field reaches its minimum in
the middle of the ring while the Cooper pair density show
different behaviors for type II and type I. The type I ring has a
positive surface energy, since the magnetic field has much larger
value at the internal boundary, the positive surface energy keeps
the Cooper pairs away. In Fig.3. the calculated super-current as a
function of radius is displayed. The current at the sample
boundary is not equal to zero which implies that the flux is not
quantized. Diamagnetic response has a negative super-current while
paramagnetic response has a positive one. Moreover, the
superconducting state can consist of a combination of the
diamagnetic and paramagnetic Meissner state, and the super-current
density goes to zero at the circle where the magnetic field
reaches a minimum. This also indicates a certain effective radius
inside the ring through which circular area the flux is exactly
quantized, which has been studied both
experimentally\cite{Pedersen:prb64} and
theoretically\cite{Yampolskii:prb}. It shall be noted that this
"combination" of superconducting state can only have diamagnetic
response in the outer part of ring and paramagnetic response in
the inner part. There is no possibility of having the response
vice versa because the super-current has to keep magnetic field
from "invading" from both the internal and the external boundary.
We have studied the case for different $\kappa$ as 0.5 and 1.0,
which represents type I and type II superconductor respectively.
There is not much difference in the behaviors of a giant vortex
state when only $\kappa$ changes, except for the value of Gibbs
free energy. We have calculated the free energy for all the cases.

The total flux through the area of the superconducting ring is
directly obtained from local magnetic field. The flux inside the
superconducting ring is found to be not necessarily quantized. For
example, the calculated flux for the giant vortex states
considered in Fig.1.(a) is around 8.7, 9.8 and 18.0 respectively.
The size of the ring is carefully chosen since it was found
experimentally that the physical properties of a mesoscopic
superconducting system depend on its size. The size of the ring we
considered here is large enough to be superconducting when more
than one flux enter. Meanwhile it is also narrow enough to
maintain the cylindrical symmetry of giant vortex states by the
boundary condition. As it has been generalized\cite{Baelus:prb61}
, the flux inside the hole of the ring is not quantized. Our
numerical results also show non-quantized solutions for the flux
inside the hole. However, the concept of winding number L in the
present work is different from that reviewed in their work. The
Gibbs free energy of the giant vortex state will not change with
L, while the flux inside the hole changes correspondingly. For
example, the flux in the hole of the combination magnetic response
in Fig.1(b). is 0.8 when $L=7$. When the winding number change to
$L=8$, the solution of GL equations is the same. The only change
is that the flux in the hole there is 1.8, which means one more
flux enters inside the hole. A transition from a state with
vorticity L to one with vorticity $L+1$ happens when a vortex,
carrying unit vorticity, enters at the outer radius, crosses the
ring, and then annihilates at the inner radius. The numerical
results are consistent with the theoretical analysis obtained in
Section II.

It is well known that the phase shift of an electron wave function
resulting in an alteration in the interference pattern of a double
slit electron diffraction experiment in the presence of a
potential magnetic field, even if the magnetic field is shielded
so that diffracted electrons do not pass through it\cite{Bohm}.
What is going to happen if only the flux inside the hole of the
superconducting ring is changed while the magnetic field is not.
If the additional flux is quantized, according to the gauge
invariant transformation Equation(13), all the parameters of the
giant vortex state will not be influenced except that the winding
number $L$ changes to $L+1$. It's also interesting to study the
case when additional flux is not quantized. Experimentally, this
additional flux could be obtained by setting a very small but long
solenoid inside the hole. The solenoid will hold in the magnetic
field inside its coils, but the uncurled vector potential will
also appear outside the coil and it will affect the
superconducting states. In Fig.4. we provide the numerical
calculation of the Gibbs free energy displayed as a function of
the additional flux. It is clearly shown that the Gibbs free
energy is a periodic function of the "field free" flux with a
period of the flux quantum $\Phi_{0}=hc/2e$. The periodicity of
Gibbs free energy with the additional flux quantum(or winding
number L) is reminiscent of Aharonov-Bohm effect, and such kind of
quantum periodicity in the free energy of superconducting pairs
has been directly observed via Little-Park experiment. If the
additional flux is not quantized, the free energy of the vortex
states will change. As a result, the magnetic field inside and
outside the superconducting ring will also change, assuming that
the additional flux will not change the magnetic field in the hole
of the ring except for the area occupied by the solenoid. In such
a way, the "invisible" vector potential $\vec{A}$ becomes
"visible". If the applied magnetic field is fixed, the additional
flux in the solenoid can even cause the free energy to become
lower. Based on our numerical results, it can be predicted that
when the superconducting state is close enough to the phase
boundary, Little-Parks like oscillations could be observed
experimentally in a rather "fat" ring for both the critical
temperature $T_{c}$ and critical applied field $H_{c}$, as the
additional flux changing continuously.

\section{Summary}

In conclusion, we have theoretically studied the giant vortex
states in terms of the non-linear GL equations for a
superconducting ring. We provide a gauge invariant transformation
and show that the solutions can be independent of the winding
number L. We also calculate the numerical results for different
magnetic responses of the superconductor, and discuss the concept
of the winding number in details. For a superconducting ring,
there are giant vortex states that hold more than one flux quantum
inside both the superconductor and the hole. The flux is found to
be not quantized, while he winding number L is still an integer,
but not necessarily to be the number of vortices anymore. The GL
equations can be fully solved without changing the value of the
winding number L. We also indicate that the periodicity of Gibbs
free energy G with additional flux in mesoscopic superconducting
ring is based on the same principle of Aharonov-Bohm effect, and
can be shown with a Little-Parks-like oscillation.

\begin{acknowledgments} This work has been
supported by the Grant from the State Education Department of
China and the Grant from Beijing Normal University. Hu Zhao also
wish to thank Professor Juelian Shen, Professor Jia-cai Nie and
Professor F. Brosens for helpful discussions.
\end{acknowledgments}


\newpage
\begin{figure}
\caption{\label{fig:desity of OP} (Color online). The amplitude of
the order parameter as a function of radius for different type of
superconductors for (a)$\kappa$=1.0 and (b)$\kappa$=0.5. The
internal boundary of the ring is at $r=10.0$, and the external
boundary at $R=12.0$. Three different magnetic responses are shown
respectively. Solid curve for the diamagnetic response,
dash-dotted curve for the combination magnetic response, and
dotted curve for the paramagnetic response.}
\end{figure}

\begin{figure}
\caption{\label{fig:magnetic field} (Color online). Magnetic field
as a function of radius for different type of superconductors
for(a)$\kappa$=1.0 and (b)$\kappa$=0.5. The internal boundary of
the ring is at $r=10.0$, and the external boundary at $R=12.0$.
Three different magnetic responses are shown respectively. Solid
curve for the diamagnetic response, dash-dotted curve for the
combination magnetic response, and dotted curve for the
paramagnetic response.}
\end{figure}

\begin{figure}
\caption{\label{fig:super-current} (Color online). Super-current
density as a function of radius for different type of
superconductors for (a)$\kappa$=1.0 and (b)$\kappa$=0.5. The
internal boundary of the ring is at $r=10.0$, and the external
boundary at $R=12.0$. Three different of magnetic response is
shown respectively. Solid curve for the diamagnetic response,
dash-dotted curve for the combination magnetic response, and
dotted curve for the paramagnetic response. Notice there appears a
zero current circle at the effective radius for the combination
magnetic response.}
\end{figure}

\begin{figure}
\caption{\label{fig:Gibbs}  Gibbs free energy as a function of
additional flux in the solenoid. The internal boundary of the ring
is at $r=10.0$, and the external boundary at $R=12.0$. GL
parameter $\kappa$ is chosen as 1.0, and $G_0$ is the Gibbs free
energy when external magnetic field vanishes.}

\end{figure}


\begin{thebibliography}{}

\bibitem{Little:prl9(9)} W.~Little and R.~D.~Parks, Phys. Rev. Lett. \textbf{9},
9 (1962).

\bibitem{London:su} F.~London, {\it Superfluids} (John Wiley \& Sons, Inc., New York, 1950), p.152.

\bibitem{Geim:apl} A.~K.~Geim, S.~V.~Dubonos, J.~G.~S.~Lok,
I.~V.~Grigorieva, J.~C.~Maan, L,~Theil~Hansen, and P.~E.~Lindelof,
Appl. Phys. Lett. \textbf{71}, 2379 (1997).

\bibitem{Moshchalkov:nature1} V.~V.~Moshchalkov, L.~Gielen, C.~Strunk, R.~Jonckheere, X.~Qiu,
C.~Van~Haesendonck, and Y.~Bruynseraede, Nature (London)
\textbf{373}, 319 (1995).

\bibitem{Geim:nature1} A.~K.~Geim, I.~V.~Grigorieva,
S.~V.~Dubonos, J.~G.~S.~Lok, J.~C.~Maan, A.~E.~Filippov, and
F.~M.~Peeters, Nature (London) \textbf{390}, 259(1997);
A.~K.~Geim, S.~V.~Dubonos, J.~G.~S.~Lok, M.~Henini, and J.~C.~Maan
\it{ibid.}\textbf{396}, 144 (1998).

\bibitem {Zhang:prb55} X.~Zhang and J.~C.~Price, Phys.
Rev. B \textbf{55}, 3128 (1997).


\bibitem{Moshchalkov:nature2} L.~F.~Chibotaru, A.~Ceulemans, V.~Bruyndoncx, and V.~V.~Moshchalkov, Nature (London) \textbf{408},
833 (2000); M.~Lange, M.~J.~Van Bael, Y.~Bruynseraede, and
V.~V.~Moshchalkov, Phys. Rev. Lett. \textbf{90}, 197006 (2003).


\bibitem{Liu:science} Y.~Liu, Y.~Zadorozhny, M.~M.~Rosario, B.~Y.~~Rock,
P.~T.~Carrigan, and H.~Wang, Science \textbf{294}
 2332 (2001).


\bibitem{Kanda:prl93} A.~Kanda, B.~J.~Baelus, F.~M.~Peeters,
K.~Kadowaki, and Y.~Ootuka, Phys. Rev. Lett. \textbf{93}, 257002
(2004).


\bibitem {Deo:prl79} P.~S.~Deo, V.~A.~Schweigert, F.~M.~Peeters,
and A.~K.~Geim, Phys. Rev. Lett. \textbf{79}, 4653 (1997).

\bibitem {Fomin:prb58} V.~M.~Fomin, V.~R.~Misko,J.~T.~Devreese and V.~V.~Moshchalkov, Phys. Rev. B \textbf{58}, 11703 (1998).

\bibitem {Deo:prl89} K.~A.~Matveev, A.~I.~Larkin, and
L.~I.~Glazman, Phys. Rev. Lett. \textbf{89}, 096802 (2002).

\bibitem {Misko:prl90} V.~R.~Misko, V.~M.~Fomin, J.~T.~Devreese and V.~V.~Moshchalkov, Phys. Rev. Lett. \textbf{90}, 147003 (2003).

\bibitem {Milosevic:prl93} M.~V.~Milosevi\'{c} and  F.~M.~Peeters, Phys. Rev.
Lett. \textbf{93}, 267006 (2004).

\bibitem {Golubovic:prb71} D.~S.~Golubovi\'{c}, M.~V.~Milosevi\'{c}, F.~M.~Peeters, and V.~V.~Moshchalkov, Phys.
Rev. B \textbf{71}, 180502(R) (2005).

\bibitem {Davidovic:prl76} D. ~Davidovi\'{c}, S.~Kumar,
D.~H.~Reich, J.~Siegel, S.~B.~Field, R.~C.~Tiberio, R.~Hey, and
K.~Ploog, Phys. Rev. Lett. \textbf{76} 815 (1996).

\bibitem {Meyers:prb68} C.~Meyers, Phys.Rev.B \textbf{68}, 104522 (2003).

\bibitem{Berger:prl75} J.~Berger and J.~Rubinstein, Phys. Rev. Lett. \textbf{75},
 320(1995); Phys. Rev. B \textbf{56}, 5124 (1997); \it{ibid.} \textbf{59}, 8896
(1999).

\bibitem {Bruyndoncx:prb60} V.~Bruyndoncx, L.~VanLook,
M.~Verschuere, and V.~V.~Moshchalkov, Phys. Rev. B \textbf{60},
10468 (1999)

\bibitem {Baelus:prb61} B.~J.~Baelus, F.~M.~Peeters, and
V.~A.~Schweigert, Phys. Rev. B \textbf{61}, 9734 (2000)

\bibitem {Pedersen:prb64} S.~Pedersen, G.~R.~Kofod,
J.~C.~Hollingbery, C.~B.~S{\o}rensen, and P.~E.~Lindelof,
Phys.Rev. B \textbf{64}, 104522 (2001).

\bibitem {Vodolazov:prb66a} D.~Y.~Vodolazov, B.~J.~Baelus, and F.~M.~Peeters,
Phys.Rev. B \textbf{66}, 054531 (2002).

\bibitem {Vodolazov:prb66b} D.~Y.~Vodolazov and F.~M.~Peeters,
Phys.Rev. B \textbf{66}, 054537 (2002).

\bibitem{Berger:prb67}J.~Berger, Phys.Rev. B \textbf{67}, 014531 (2003).

\bibitem {Vodolazov:prb67} D.~Y.~Vodolazov, F.~M.~Peeters,
S.~V.~Dubonos, and A.~K.~Geim, Phys.Rev. B \textbf{67}, 054506
(2003).

\bibitem{Morelle:prb70} M.~Morelle, D.~S.~Golubovi\'{c}, and
V.~V.~Moshchalkov, Phys. Rev. B \textbf{70}, 144528 (2004)

\bibitem {Bourgeois: prl94} O.~Bourgeois, S.~E.~Skipetrov, F.~Ong,
and J.~Chaussy, Phys. Rev. Lett. \textbf{94}, 057007 (2005).

\bibitem{Stenuit:physica00} G.~Stenuit, J.~Govaerts, D.~ Bertrand, and O. van der Aa, Physica C
\textbf{332}, 277 (2000).

\bibitem {Milosevic:prb02} M.~V.~Milosevi\'{c}, S.~V.~Yampolskii, and  F.~M.~Peeters
Phys.Rev. B \textbf{66}, 024515 (2002)



\bibitem{Press:book1} W. H. Press, B. P. Flannery, S. A. Teukoslky, and W. T.
Vetterling, \textit{Numerical Recipes: The Art of Scientific
Computing}, Cambridge University Press, 1992.

\bibitem{Yampolskii:prb} S.~V.~Yampolskii,
F.~M.~Peeters,B.~J.~Baelus,and H.~J.~Fink,Phys. Rev. B
\textbf{64}, 052504 (2001)

\bibitem{Bohm} Y.~Aharonov and D.~Bohm, Phys.
Rev. \textbf{115}, 485 (1959).


\end{thebibliography}
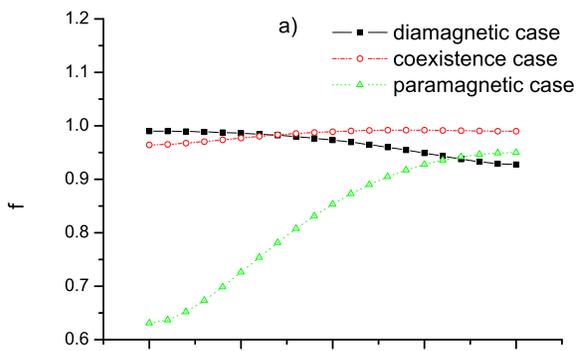
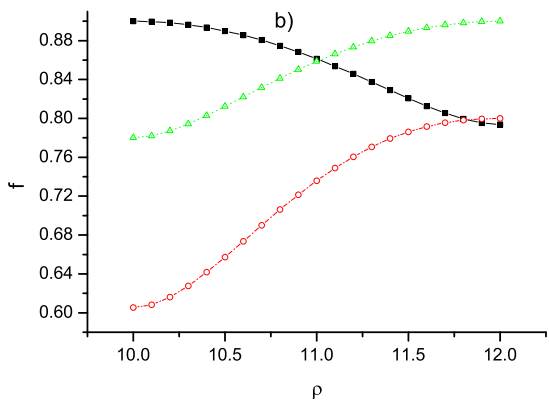
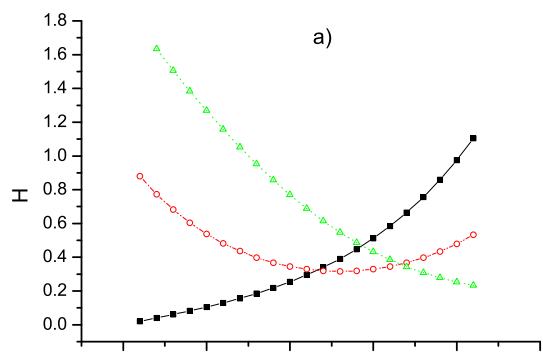
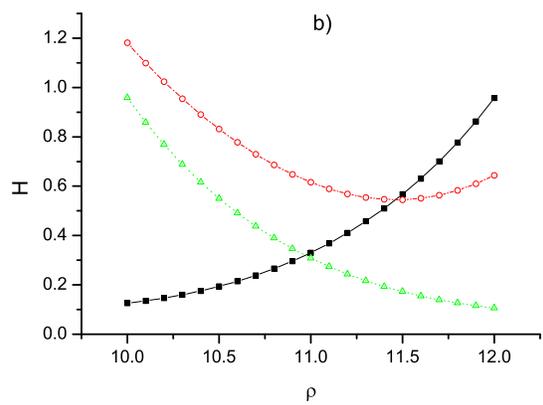
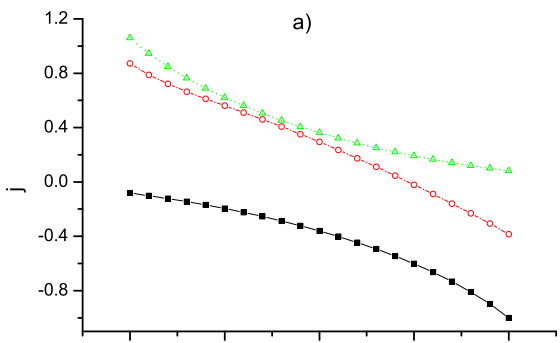
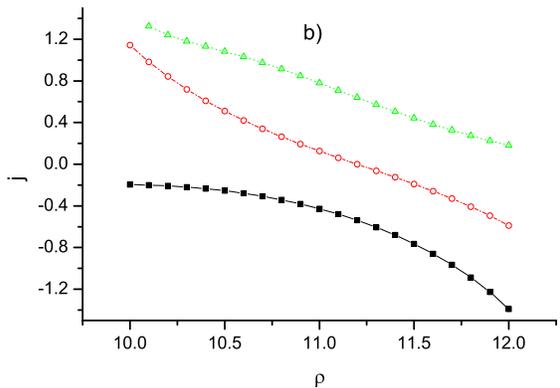
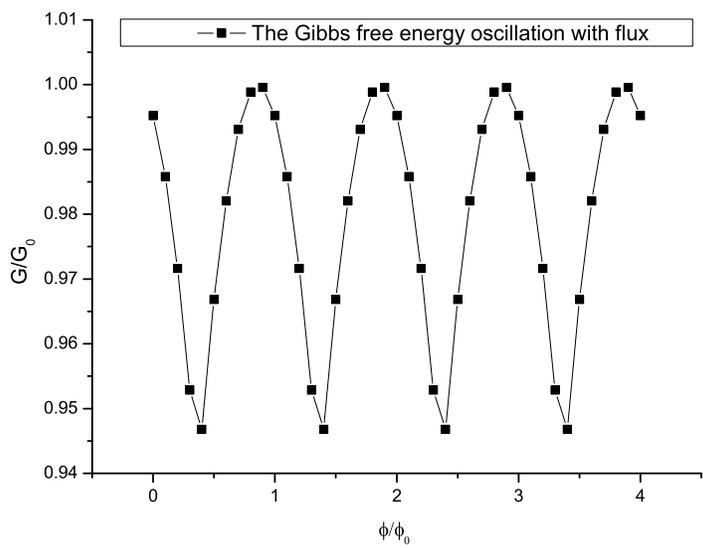
\end{document}